\documentclass[conference]{IEEEtran}

\IEEEoverridecommandlockouts
\usepackage{cite}
\usepackage{amsmath,amssymb,amsfonts}
\usepackage{textcomp}

\usepackage{graphicx}
\usepackage{caption}
\usepackage{placeins}
\usepackage{hyperref}
\usepackage{subcaption}
\usepackage{enumitem}
\usepackage{algorithm}
\usepackage{algpseudocode}
\usepackage{xcolor}
\usepackage{siunitx}
\usepackage{amsmath}

\def\BibTeX{{\rm B\kern-.05em{\sc i\kern-.025em b}\kern-.08em
    T\kern-.1667em\lower.7ex\hbox{E}\kern-.125emX}}

\begin{document}

\newcommand{\sulf}{SO\textsubscript{2}F} 
\newcommand{\DD}[1]{\textcolor{red}{[DD: #1]}}

\title{Circuit Depth Reduction for Executable Hamiltonian Dynamics of Covalent Inhibitor Reactivity on Quantum Hardware}
\author{

\IEEEauthorblockN{Marek Kowalik}
\IEEEauthorblockA{\textit{Capgemini Quantum Lab}\\
Paris, France\\
ORCID: 0009-0007-8484-1315}
\vspace{-0.6em}

\and
\IEEEauthorblockN{Sam Genway}
\IEEEauthorblockA{\textit{Capgemini Quantum Lab}\\
Paris, France\\
ORCID: 0009-0006-9143-1691}
\vspace{-0.6em}

\and
\IEEEauthorblockN{Vedangi Pathak}
\IEEEauthorblockA{\textit{IBM T. J. Watson Research Center}\\
Yorktown Heights, NY, USA\\
ORCID: 0000-0001-6335-6539}
\vspace{-0.6em}

\and
\IEEEauthorblockN{Mykola Maksymenko}
\IEEEauthorblockA{\textit{Haiqu Inc.}\\
San Francisco, CA, USA\\
ORCID: 0000-0003-0685-5790}
\vspace{-0.6em}

\and
\IEEEauthorblockN{Simon Martiel}
\IEEEauthorblockA{\textit{IBM Quantum}\\
\textit{IBM France Lab}\\
Orsay, France\\
ORCID: 0000-0001-5624-2955}
\vspace{-0.6em}

\and
\IEEEauthorblockN{Hamed Mohammadbagherpoor}
\IEEEauthorblockA{\textit{IBM T. J. Watson Research Center}\\
Yorktown Heights, NY, USA\\
ORCID: 0000-0001-9707-5305}
\vspace{-0.6em}

\and
\IEEEauthorblockN{Richard Padbury}
\IEEEauthorblockA{\textit{IBM T. J. Watson Research Center}\\
Yorktown Heights, NY, USA\\
ORCID: 0009-0003-9772-0521}
\vspace{-0.6em}

\and
\IEEEauthorblockN{Vladyslav Los}
\IEEEauthorblockA{\textit{Haiqu Inc.}\\
San Francisco, CA, USA\\
ORCID: 0009-0007-4781-6938}
\vspace{-0.6em}

\and
\IEEEauthorblockN{Oleksa Hryniv}
\IEEEauthorblockA{\textit{Haiqu Inc.}\\
San Francisco, CA, USA\\
ORCID: 0009-0009-6414-4912}
\vspace{-0.6em}

\and
\IEEEauthorblockN{Peter Pog\'{a}ny}
\IEEEauthorblockA{\textit{GSK Medicines Research Centre}\\
Stevenage, United Kingdom\\
ORCID: 0000-0003-3536-0746
}
\vspace{-0.6em}

\and
\IEEEauthorblockN{Phalgun Lolur$^*$}
\IEEEauthorblockA{\textit{Capgemini Quantum Lab}\\
Paris, France\\
$^*$ Corresponding author: phalgun.lolur@capgemini.com}
\vspace{-0.6em}
}


\maketitle

\begin{figure}[htbp]
    \centerline{\includegraphics[width=0.75\columnwidth]{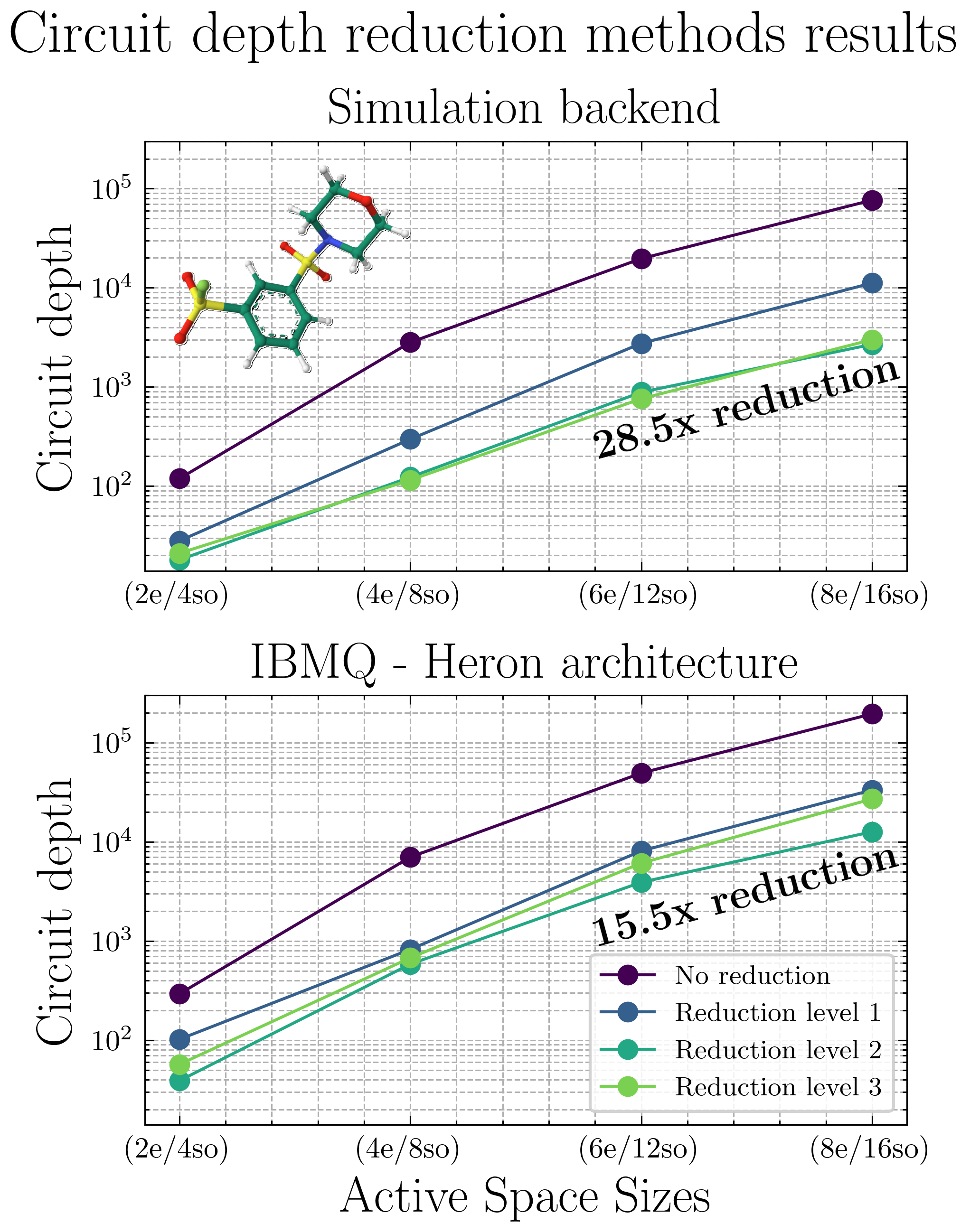}}
    \caption*{\textit{Mean circuit depth after 50 trials from an exemplary molecule for 1 Trotter step. Top figure shows transpilation on all-to-all connectivity with $\textit{Clifford}+R_z$ set of gates and bottom figure IBM Heron R2 architecture. The best reduction from all the methods and active spaces per each backend is annotated.}}
    \label{abstract_figure} 
\end{figure}

\begin{abstract}
Quantum chemistry applications in the noisy intermediate-scale quantum era require end-to-end approaches that balance algorithmic fidelity with practical executability on existing hardware. We present an end-to-end Hamiltonian dynamics case study for predicting the reactivity of pharmaceutically relevant covalent inhibitors containing sulfonyl fluoride warheads, using a quantum-centric data-driven research and development framework that combines Hamiltonian time evolution with classical machine learning. To make such simulations executable on current quantum processors, we introduce a systematic circuit reduction strategy based on Hamiltonian term truncation with observable error bounds, Clifford Decomposition and Transformation, and hardware-aware transpilation. Across representative molecular fragments, this approach achieves circuit depth reductions of up to 28.5× under all-to-all connectivity assumptions and up to 15.5× on IBM Heron-class architectures. For an eight-qubit Hamiltonian dynamics simulation, a transpiled instruction set architecture (ISA) circuit depth of 1330 is rendered executable through middleware-enabled circuit decomposition, enabling the execution of sub-circuits with depths up to 371 and containing up to 216 two-qubit gates on real hardware. We evaluate the impact of circuit reduction on downstream reactivity prediction accuracy and show that chemically meaningful predictions can be retained despite aggressive circuit simplifications, clarifying the trade-offs that govern practical quantum chemistry workflows on near-term quantum systems. 
\end{abstract}




\FloatBarrier

\section{Introduction}
\label{sec:introduction}

Quantum chemistry simulation is seen as one of the most promising applications of quantum computing \cite{Alexeev2024}. Specifically, electronic structure Hamiltonian simulations play a central role in extracting chemistry insights from commercially relevant materials \cite{Chan2024}. In the pharmaceutical industry, these simulations are particularly valuable for drug discovery, where accurate predictions of molecular properties can significantly accelerate development timelines and reduce costs.

In the domain of drug discovery, \textit{in silico} methods encompassing machine learning and molecular simulations have gained prominence over four decades, driven by improved computational power and expanded experimental datasets. However, one of the central challenges in designing targeted covalent drugs—an increasingly important class of therapeutics—is the accurate prediction of warhead reactivity, which critically influences drug potency and selectivity \cite{McAulay2022}. Unlike traditional small-molecule inhibitors, covalent drugs \cite{Gilbert2023,Montgomery2023} operate through a two-step mechanism: initial reversible protein binding followed by covalent bond formation between the drug's warhead and a target amino acid residue. This field has experienced considerable recent growth \cite{Bhole2024,Boike2022,Singh2022}, driven by advances in chemoproteomic assays enabling proteome-wide studies.

Current computational approaches for predicting drug reactivity face significant limitations. For sulfonyl fluoride (\sulf) warheads, which target non-thiol containing amino acids, density functional theory (DFT) calculations of LUMO energies have shown success \cite{Gilbert2023}. However, other warheads require more complex transition state calculations \cite{Lonsdale2017}, making the computational cost prohibitive for large-scale screening. These limitations motivate the exploration of quantum computing solutions.

In the current noisy intermediate-scale quantum (NISQ) era, researchers must frame commercially valuable problems into scalable solutions that leverage existing quantum hardware \cite{Alexeev2024}. While initial quantum chemistry implementations focused on small-scale systems \cite{Montgomery2023}, recent advances have enabled more accurate electronic structure predictions for larger quantum systems \cite{Shajan2024} through enhanced error mitigation techniques \cite{Shajan2024} and hybrid quantum-classical algorithms \cite{Kanno2023}. Notable achievements include Sample-based Quantum Diagonalization calculations of ground and excited states for iron-sulfur clusters with active spaces up to (54e, 36o) \cite{Robledo-Moreno2024, Shajan2024, Barison2024}.

One promising approach is the Quantum-Centric Data-Driven Research and Development (QDDRD) framework\cite{Montgomery2023}, which combines quantum simulations with machine learning (ML) to generate "quantum fingerprints" for predictive modeling. This methodology transforms challenging excited-states calculations into ML predictions based on Hamiltonian simulations. However, simulating dynamics—essential for understanding molecular interactions and reactions—presents additional challenges due to the complexity and extended simulation times required, resulting in prohibitively large circuit depths.

As quantum processing units (QPUs) continue to advance with reduced error rates, significant opportunities exist to optimize quantum algorithms\cite{Berry2024, Ostmeyer2023} through circuit depth reduction\cite{Mukhopadhyay2023, Harrow2024} and decreased sampling overhead\cite{Lee2024, VanDam2024, Liao2023, Alexeev2024}. In this work, we address these challenges by developing a quantum circuit reduction framework for Hamiltonian simulation. We achieve up to 15.5-fold circuit depth reduction while maintaining predictive accuracy, demonstrating a practical approach to scaling quantum chemistry calculations on NISQ devices. Our results can be extrapolated to larger active spaces through hardware improvements and algorithmic refinements.

\section{Background}
\label{sec:background}

Hamiltonian simulation requires careful consideration in selecting appropriate methods and algorithms for implementation on current quantum hardware. Among the available approaches for real-time evolution simulation, several methods stand out: product formulas (PF), quantum walks\cite{berry2012}, fractional-query simulations\cite{berry2014}, Taylor series (TS)\cite{berry2015} and quantum signal processing (QSP)\cite{Berry2024}. While QSP may be theoretically optimal for certain Hamiltonians in terms of gate complexity \cite{Low2017, Childs2018}, PF presents distinct advantages in the NISQ era, particularly for smaller active spaces. PF's appeal lies in its general simplicity, avoiding large unitary control gates, and ability to generate relatively shallow circuits while maintaining strong approximations through first-order formulas with few Trotter steps. In contrast, methods like QSP typically require deep circuits that are impractical without quantum error correction. Furthermore, PF demonstrates better scalability than TS and QSP when applying accuracy empirical bounds on Trotter steps \cite{Reiher2017}, as demonstrated in studies of 1D-spin lattice systems \cite{Childs2018}. To optimize experiments for larger active space sizes, the circuits generated by PF can be further refined through additional methods including Hamiltonian terms truncation, Clifford Decomposition and Transformation (CDAT)\cite{Gujarati2023}, and qubit tapering\cite{Bravyi2017, Setia2020}. While previous implementations of electronic structure Hamiltonian simulations on real quantum hardware have been limited by circuit depth and width constraints, our work pushes these boundaries significantly. Through careful optimization of the PF approach combined with circuit reduction techniques, we demonstrate simulations with a transpiled ISA depth of 1330 on 8 qubits, marking an advance in practical quantum chemistry calculations on NISQ devices.

\subsection{QDDRD framework}
\label{subsec:QDDRD_framework}

\begin{figure}[htbp]
\centerline{\includegraphics[width=\linewidth]{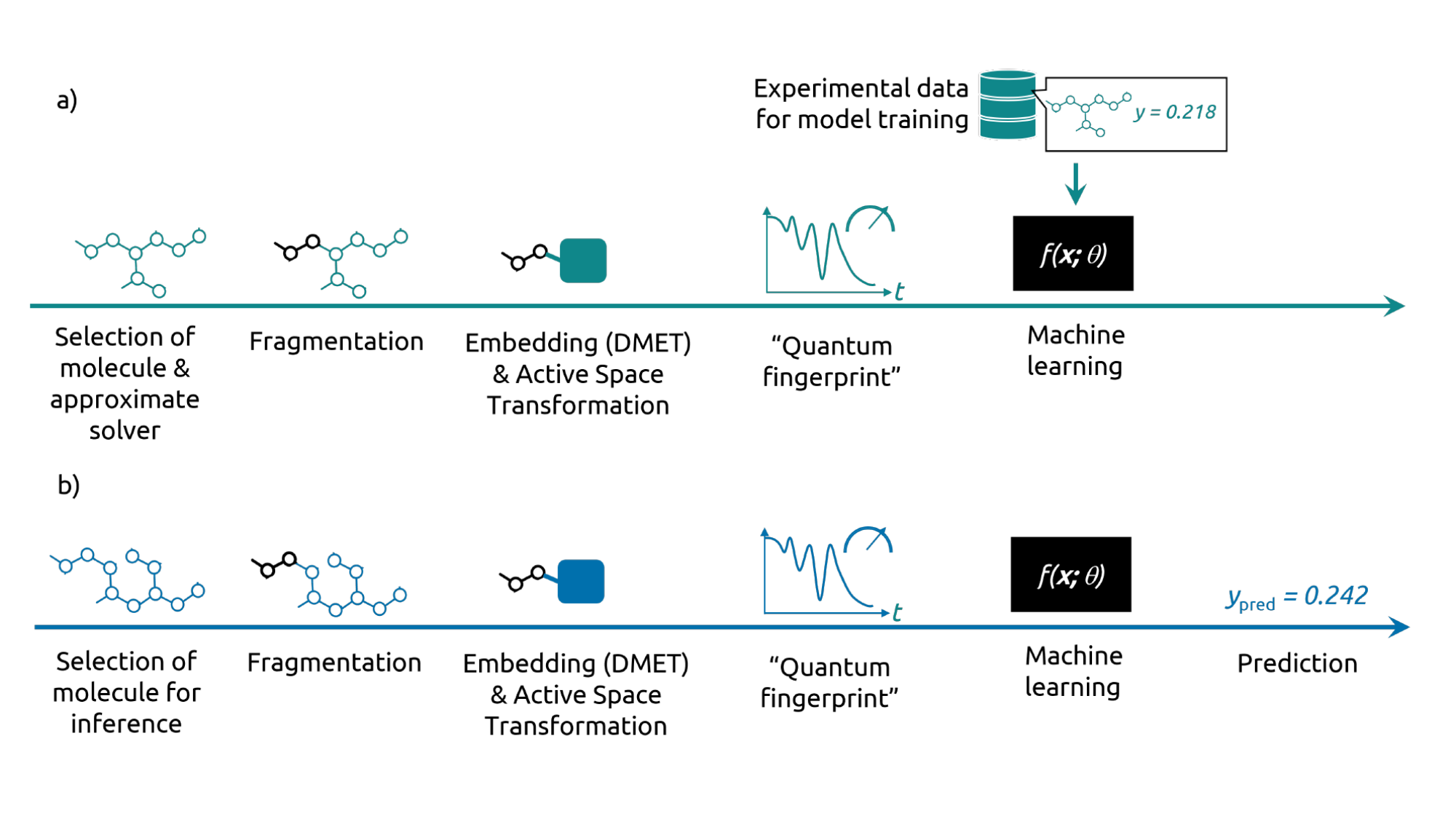}}
\caption{High-level representation of the data-driven pipeline used to predict molecule reactivities using quantum computers and ML. Training stage a) is conducted using experimental reactivity data or data calculated classically using DFT, as a ground truth for the model. Subsequently, the trained model is used in the inference stage b) to predict new molecular reactivities. Figure taken exactly from \cite{Montgomery2023} with permission.}
\label{fig:QDDRD pipeline}
\end{figure}

The quantum chemistry problem highlighted in this work is the reactivity prediction of a series of targeted covalent drug inhibitors using the `quantum features' as described in \cite{Montgomery2023}. Following on from this previous work, we perform predictions using the data-driven pipeline shown in Figure \ref{fig:QDDRD pipeline} that leverages gate-based quantum computers to calculate quantum features.

In this approach, a series of molecules with chemical similarity is explored using a molecular fragment relevant to the problem to reduce the complexity of the problem. Once the fragment of interest has been defined, suitable embedding methods and active space selection are performed to obtain an effective Hamiltonian that describes the fragment and its entanglement with the rest of the molecule (and eventually the rest of the environment around the molecule e.g. water). The effective Hamiltonian is constructed to obtain insights into the dynamics of the system, from which the input features for the machine learning model may be derived. In our case, we select expectation values of data-driven temporal observables, measured at the certain time-evolved states of our system, called `\textit{quantum fingerprints}'. To perform this task using quantum computers, the Hamiltonian dynamics of the system will be simulated as shown in Section~\ref{sec:background}. Finally, the machine learning model can be trained to predict an interesting objective by using the training (and testing) dataset of molecules with experimental measurement values or simulated ground truths from classical computation.

Worth mentioning is the fact, that beyond reactivity prediction, Hamiltonian simulation has also shown promise in providing quantum fingerprints that enable effective clustering of molecules based on their electronic structure properties \cite{Montgomery2023}, offering practical value for drug discovery applications even in the NISQ era.

\section{Methodology}
\label{sec:methodology}

This section describes the end‑to‑end methodology used to execute Hamiltonian dynamics simulations for covalent inhibitor reactivity prediction on current quantum hardware. The approach integrates problem formulation, quantum circuit construction, circuit depth reduction, hardware‑aware compilation, and execution management within a hybrid quantum–classical workflow. The emphasis throughout is on execution feasibility rather than algorithmic ideality, with each methodological choice evaluated in terms of its impact on circuit depth, noise exposure, and downstream predictive performance. 

\FloatBarrier
\noindent
\begin{minipage}{\linewidth}
    \captionsetup{type=figure} 
    \subcaptionbox{}{\includegraphics[width=0.30\linewidth]{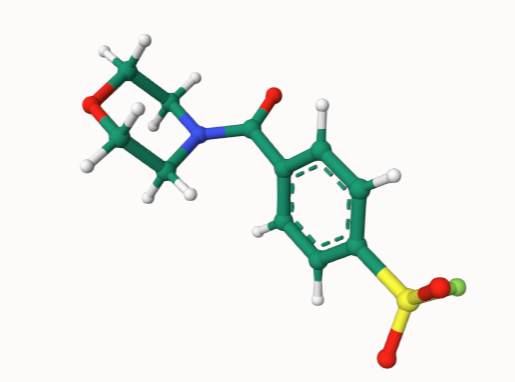}}
    \quad
    \subcaptionbox{}{\includegraphics[width=0.31\linewidth]{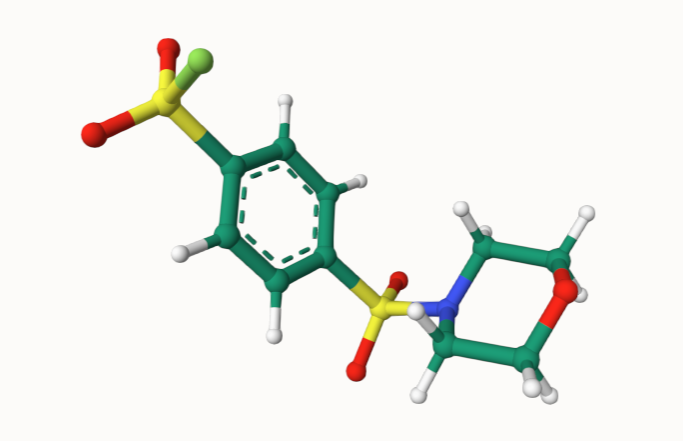}}\quad
    \subcaptionbox{}{\includegraphics[width=0.3\linewidth]{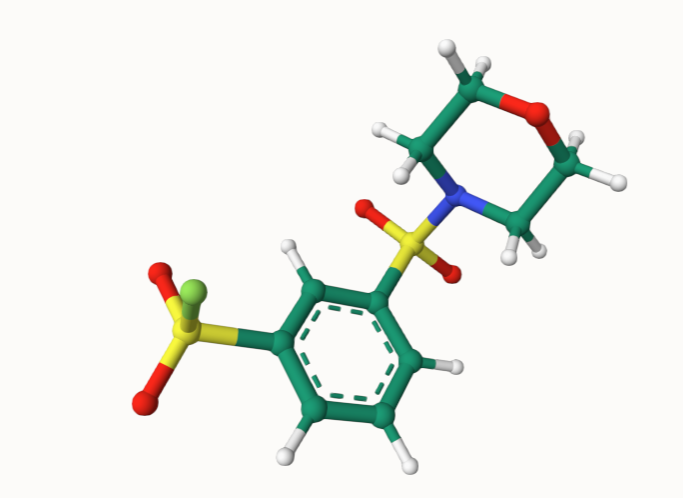}}
    \caption{2D diagram structures of 3 exemplary molecules from the dataset. Respectively a) 4-(Morpholine-4-carbonyl)benzene-1-sulfonyl fluoride, b) 4-(Morpholine-4-sulfonyl)benzene-1-sulfonyl fluoride, c) 3-Morpholin-4-ylsulfonylbenzenesulfonyl fluoride.}
    \label{fig:dataset_3_mols_example_diagrams}
\end{minipage}
\vspace{\baselineskip}

Our approach combines established quantum chemistry simulation techniques with novel circuit optimization methods to enable larger-scale Hamiltonian simulations on current quantum hardware. Following the methodology outlined in \cite{Montgomery2023}, we extend their framework through several key enhancements designed to handle larger active spaces on real quantum processing units (QPUs), ultimately achieving electronic structure Hamiltonian simulations with a transpiled ISA depth of 1330 on 8 qubits; after decomposition within the middleware routine, these circuits were executed as sub-circuits compatible with current quantum hardware. The methodology begins with a characterization of our pharmaceutical dataset and ground truth calculations (Subsection~\ref{subsec:use_case_details}), followed by implementation of a comprehensive quantum experiments pipeline that transforms qubit-mapped Hamiltonians into feature vectors for reactivity prediction in the Subsection~\ref{subsec:quantum_experiments_pipeline}. We then describe our transpilation strategies for instruction set architecture preparation. The integration of Haiqu middleware streamlines our quantum workflows through enhanced circuit analysis, hardware-aware compilation, and error mitigation capabilities (Subsection~\ref{subsec:middleware_application}). Finally, we present our approach to quantum error mitigation and suppression, emphasizing methods with manageable overhead suitable for integration with middleware (Subsection~\ref{subsec:error_mitigation_methods}). Together, these components form a comprehensive methodology for executing end‑to‑end hybrid Hamiltonian simulations workflow on current quantum hardware while maintaining prediction accuracy.

\subsection{Use case details}
\label{subsec:use_case_details}

The dataset includes a series of targeted covalent drugs with the sulfonyl fluoride ($SO_2F$) reactive group (known as a ‘warhead’), which is also the fragment of interest. Raw data contain 274 molecules with 276 unique 3D structures. In this work, we select 8 molecules from the dataset of the original paper \cite{Montgomery2023} while the remaining molecules were taken from \href{https://pubchem.ncbi.nlm.nih.gov/}{PubChem} database with geometrical structure refinement using force-field optimization with MMFF94 variant from \href{https://www.rdkit.org}{RDKit}\cite{rdkit}. The split for cross-validation was 80\% for the training dataset and 20\% for the test dataset. The chemical structure of 3 representative molecules is shown in Figure \ref{fig:dataset_3_mols_example_diagrams}.

Ground truth reactivities were simulated with DFT as a linear regression between the measured rate constant (or half-life) and the LUMO energy \cite{Gilbert2023,Montgomery2023}. To create the electronic structure Hamiltonians, LUMO energies calculated with B3LYP-D3 functional on the \textit{6-31+G**} basis set were used. The rationale behind not using directly the experimental values is to give a proof of concept for reproducing values coming from quantum mechanical calculations on classical computers. The embedding is done using Density Matrix Embedding Theory (DMET) along with the active space reduction described in \cite{Montgomery2023}.  The initial state of the simulated fragments will be Hartree-Fock initial state, since for this dataset and this problem it yields the best ML model performance in comparison to other initial states \cite{Montgomery2023}. As the observable yielding the quantum fingerprint, the temporal observable will be taken, that is defined as:
\begin{equation}
F(t)=\sum_{r,s}{h_{rs}^{eff}\rho_{rs}(t)}\label{eq:temporal_observable}
\end{equation}
Where $\rho_{rs}(t)=\langle\psi(t)|{\hat{a}}_s^\dagger{\hat{\ a}}_r|\psi(t)\rangle$ is the one-body density matrix of the embedded fragment, where $r$ and $s$ are the localized orbitals in the chosen fragment and the $h_{rs}^{eff}$ are the effective one-electron integrals from the embedded effective Hamiltonian. This observable yields a quadratic number of terms, so a quadratic number of Pauli measurement operators from the active space size. Based on this observable expectation value, the feature vector will be created as the sampled observable trajectory within the time range $t\in[0,\ 14]\ t/t_h$ with the sampling interval $\Delta t=0.5\ $ $t/t_h$. Finally, the objective of the machine learning model is the molecule reactivity. The ground truth will be the reactivity simulated with classical methods as described in the previous section. As the model architecture, the Partial Least Squares Regression (PLS) will be used. 

\subsection{Quantum experiments pipeline}
\label{subsec:quantum_experiments_pipeline}

This section describes the procedure of obtaining quantum fingerprints in the use case, starting with the qubit-mapped Hamiltonian and the initial Hartree-Fock state and ending with the input feature vector for the reactivity prediction model. 

The initial state, the Hamiltonian, and the observables in the second quantized fermionic form need to be qubit mapped. For the scope of this paper Jordan-Wigner was chosen. More advanced methods are pointed out in the Discussion Section~\ref{sec:discussion}.

All simulations will be run with the 1\textsuperscript{st} order PF decomposition with 1 Trotter step, since the main goal is to get the Hamiltonian simulation method with shallowest circuits feasible to run on real QPUs for larger active space sizes, no matter the algorithmic approximation done by the method. The idea is to reduce the circuits yielded by the method until it is feasible to run on a real QPU. That allows us to evaluate the accuracy of the predictive models with the current QPUs capability. Based on that, further work on increasing the accuracy of the solution (either by using further circuit reduction methods or running experiments on QPUs of higher fidelities) can be done. Also, circuit depths obtained for this simplest case may be easily extrapolated on more Trotter steps and higher-order PFs. That will be helpful for resource estimation of accurate approximations.
The accuracy of our approximation will be assessed in the Results Section~\ref{subsec:hamiltonian_terms_truncation_results} by evaluating the predictive model accuracy with the Root Mean Squared Error (RMSE) metric. For the ground truth values of the predicted reactivity, both for model training and inference, classically calculated reactivity as described in the Subsection above will be used.

\begin{figure}[h!]
    \centering
    \includegraphics[width=\linewidth]{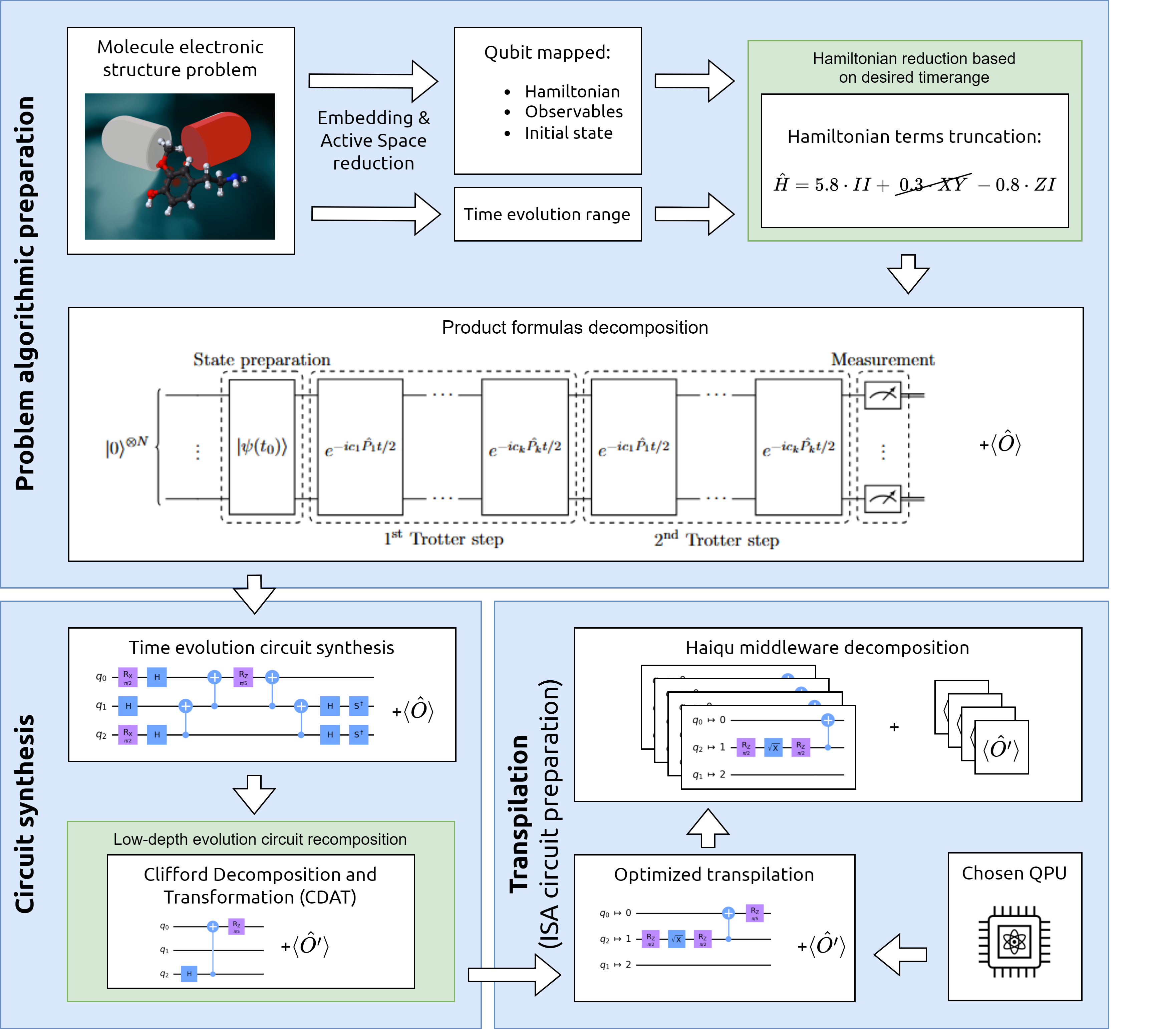}
    \caption{Flowchart of the optimized real-time evolution simulation execution on quantum computers consisting of three parts - Problem algorithm preparation, Circuit synthesis, and Transpilation. Blue rectangles mark the steps of the time-evolution circuit preparation.}
    \label{fig:methodology_flowchart}
\end{figure}

In regards to the extension of original work, this paper extends previous work with the enhancements allowing to run active space sizes larger than the smallest one (2e, 4so) on real QPUs. Those improvements can be applied on four levels, reflecting also sequentially how they are executed during real-time evolution simulation routine:
\begin{enumerate}[itemsep=0pt,parsep=0pt]
    \item Problem algorithmic preparation - initial state, Hamiltonian and observables preparation, qubit mapping, and PF decomposition
    \item Simulation circuit synthesis - circuit construction from the initial state and Hamiltonian terms exponentiation
    \item ISA circuit preparation - synthesized circuit transpilation on the desired backend
    \item Quantum error mitigation and suppression - specification of the error mitigation and suppression protocols and running the experiments using these protocols 

\end{enumerate}
On each level, improvements may be generic (problem-agnostic) and problem-aware, leveraging information about the problem. The workflow combines existing techniques with methodological choices introduced here for this Hamiltonian-dynamics use case:
\begin{enumerate}[itemsep=0pt,parsep=0pt]
    \item Problem algorithmic preparation - truncation of Hamiltonian terms (described in Appendix~A)
    \item Simulation circuit synthesis - Clifford Decomposition and Transformation (CDAT) \cite{Gujarati2023}
    \item Transpilation (ISA circuit preparation) - transpilation techniques particularly focused on gates cancellation, effective assigning virtual qubits to physical ones generat ing the least amount of 2-qubit gates (Subsection~\ref{subsec:ISA_circuit_preparation}.)
    \item Error mitigation and suppression - build-in protocols in Qiskit\cite{Qiskit} (Subsection~\ref{subsec:error_mitigation_methods})
\end{enumerate}
On top of that, Haiqu middleware is incorporated (Subsection~\ref{subsec:middleware_application}) and it covers levels 2, 3, and 4. The technique from the first level is described in Appendix~A, while the details of the last two levels' details are provided in this subsection. A graphical composition of the techniques used can be found in Figure \ref{fig:methodology_flowchart}. The circuit depths obtained after straightforward exponentiation of the qubit mapped Hamiltonians to the ${\text{\it{Clifford}}+R_z}$ set of gates, for a range of active spaces, will be the point of reference for circuit reductions in the presented methods. 

\noindent

The optimization starts from the qubit-mapped Hamiltonian terms. Terms with small coefficients cannot significantly influence the measured observable expectation value, since only a narrow initial range for the time evolution is sampled ( $t\in[0,\ 14]\ t/t_h$).
Effective qubit Hamiltonians obtained from the used dataset for checked ranges of active space sizes (2e,4so)-(22e, 44so) have coefficients of terms spanning orders of magnitudes ($10^{-1}-10^{-7}$) that allow to safely exclude most of the terms for the simulation. 
The exact truncation algorithm that was used in this paper is based on the maximum expectation value's change $\epsilon_{d\langle\hat{A}\rangle}$ induced by removing the terms out of the Hamiltonian. It is introduced in ~Appendix~A. The next step is the optimization on the circuit synthesis level. There, the CDAT transformation is applied to further reduce the circuit depth, by effectively getting rid of the majority of the Clifford gates. Results for simulation circuit reduction for the combination of CDAT and truncation based on the observable error for max. range of simulation time are presented in the Results Section~\ref{subsec:circuits_reduction_results}.

\subsection{Instruction Set Architecture (ISA) circuit preparation}
\label{subsec:ISA_circuit_preparation}

The translation of abstract quantum circuits to hardware-executable instructions requires careful optimization. We utilized Qiskit's transpiler (version 1.2) with optimization level 2, leveraging advanced features including swap gate minimization, inverse gate cancellation, and optimized qubit routing through the `sabre' method.

For comparison we also tried the algorithm presented in \cite{Rustiq2024}, called \textit{Rustiq}, in order to generate the evolution circuit, before mapping it to the device. This algorithm solves the problem of implementing a given sequence of multi-qubit Pauli rotations using only single- and two-qubit gates. It proceeds by scoring each and every possible single-CNOT piece of the circuit (called `chunks') by conjugating all the remaining Pauli rotations through it. The score of a chunk will roughly correspond to how much progress was made in reducing the weights of all remaining rotations, with an emphasis on the first rotations in the list.
The algorithm greedily builds a circuit by picking the highest scoring chunk, conjugating all rotations through it, injecting weight one rotations as single-qubit gate, and carrying on. By considering the full sequence of rotations when scoring each chunk, this algorithm is able to exploit global symmetries of the input to greatly reduce the overall gate count and depth.
This approach can improve the depth of Hamiltonian simulation circuits by a factor of 10 on standard chemistry benchmarks \cite{Rustiq2024}.

\subsection{Middleware implementation}
\label{subsec:middleware_application}

To handle the challenges of executing deep quantum circuits on real QPUs, we employed Haiqu middleware, which provides end-to-end optimization of the quantum computing workflow, from circuit analysis and hardware-aware compilation to error mitigation and execution management. Haiqu middleware targets each aspect of the quantum computing workflow to streamline it: analysis of the circuits' properties, particularly in relation to the intended hardware, custom hardware-aware compilation, error mitigation (EM), circuit execution and post-processing. The system provides both improved versions of some of the individual components as well as the capacity to orchestrate stacks of these techniques and combine them with the Optimized Execution strategy. 

\begin{figure}
    \centering
    \includegraphics[width=1.01\linewidth]{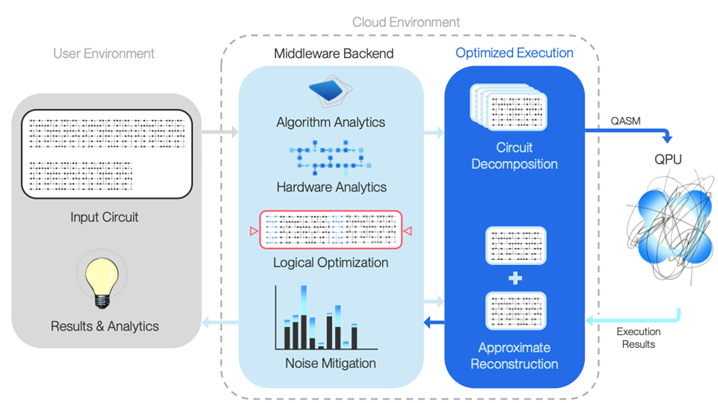}
    \caption{Workflow of the Optimized Execution strategy developed by Haiqu. It is a strategy for executing multiple modified algorithm sub-blocks in a sequential or parallel fashion and then reconstructing the final or intermediate quantum state. It is distinct from circuit partitioning and synthesis strategies and combines aspects of standard compilation and error mitigation techniques.}
    \label{fig:haiqu_flowchart}
\end{figure}

The proprietary Optimized Execution (OE) strategy is based on executing, in a sequential or parallel fashion, multiple modified algorithm sub-blocks and subsequently reconstructing the final or intermediate quantum state. It superficially resembles, but is distinct from, circuit partitioning\cite{Mitarai2021} and synthesis strategies e.g. \cite{Patel2003}, and combines aspects of both standard compilation and error suppression techniques. Each individually executed sub-block is ensured to fit the practical performance constraints of the QPU, minimizing the accumulated errors, and allowing to reproduce the results of running a quantum algorithm whose length would naively place it beyond the performance range of the device. As such an execution strategy is not directly a compilation approach nor a noise mitigation technique, it can be combined with other techniques to further improve the results. The high-level flow diagram is shown at Figure \ref{fig:haiqu_flowchart}. 

\FloatBarrier

\subsection{Error mitigation methods}
\label{subsec:error_mitigation_methods}

To mitigate the impact of hardware noise, a combination of low‑overhead error mitigation and suppression techniques is applied, including  Twirled readout error extinction\cite{van_den_Berg_2022} for readout error mitigation, Pauli twirling\cite{Wallman2016}, and dynamical decoupling\cite{Ezzell2023}. These methods are selected to be compatible with the middleware execution strategy and to avoid excessive sampling overhead. All quantum experiments are executed within a single calibration window to minimize systematic drift. 



\section{Computational Details}
\label{sec:computational_details}

For the software, the classical computing work was done in Python with Scikit-learn\cite{scikit-learn},  PySCF\cite{Sun2018}, RDKit\cite{rdkit}, Vayesta\cite{Nusspickel2023} and Qiskit (particularly Qiskit Nature) packages\cite{Qiskit, QiskitAddonOBP}. All classical computation was run locally on a laptop with Intel i7 9\textsuperscript{th} gen CPU 2.6GHz, NVIDIA Quadro T1000, and 32GB of RAM. 
Quantum computing parts of this paper were set up and run using native Qiskit 1.2 and its extensions in Python, both for reference, ideal simulations, and real backend experiments. 
Ideal simulations were run on Qiskit Aer simulator. All QPU estimations were run on IBM's superconducting qubits backends with the newest IBM's Heron R2 architecture\cite{AbuGhanem2024} in the batch mode. Each time point was run during the same calibration cycle with a total number of 8192 shots, 64 Pauli twirls (128 shots per twirl), 16 T-REx randomizations (100 shots per randomization), and dynamical decoupling with `$X_pX_m$' type of sequence.
For the Haiqu middleware runs, all details are provided in Appendix~B.
\FloatBarrier

\section{Results}
\label{sec:results}

Our results demonstrate a two-fold contribution: first, a Hamiltonian-simulation reduction and validation workflow combining observable-guided truncation, CDAT optimization, and hardware-aware transpilation, which yields a transpiled ISA depth of 1330 for the eight-qubit demonstration circuit; second, a hardware execution demonstration in which Haiqu middleware provides the decomposition layer needed to run the resulting circuits on current QPUs and recover trajectories close to the ideal simulation. This achievement represents a significant advance in the practical implementation of quantum chemistry calculations on NISQ devices, enabled by our circuit reduction approach combined with middleware-enabled execution.

\subsection{Hamiltonian terms truncation results}
\label{subsec:hamiltonian_terms_truncation_results}

\begin{figure}
    \centering
    \includegraphics[width=1\linewidth]{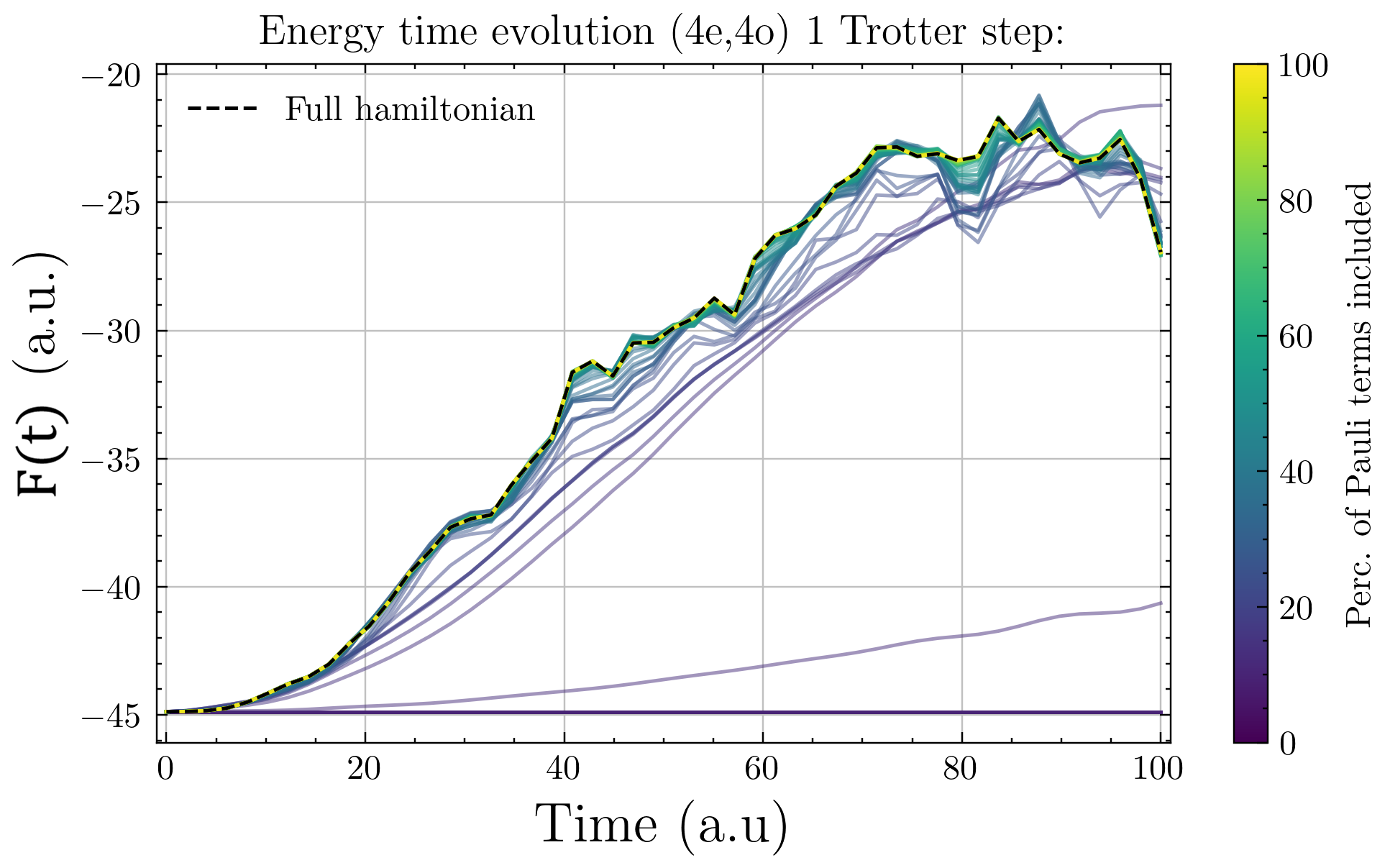}
    \caption{Temporal observable $F(t)$ emulated without any noise and without sampling over a range of terms left in the Hamiltonian for an exemplary molecule of the active space with 8 spin-orbitals.}
    \label{fig:time_evo_example_from_terms_included}
\end{figure}

\begin{figure}
    \centering
    \includegraphics[width=0.99\linewidth]{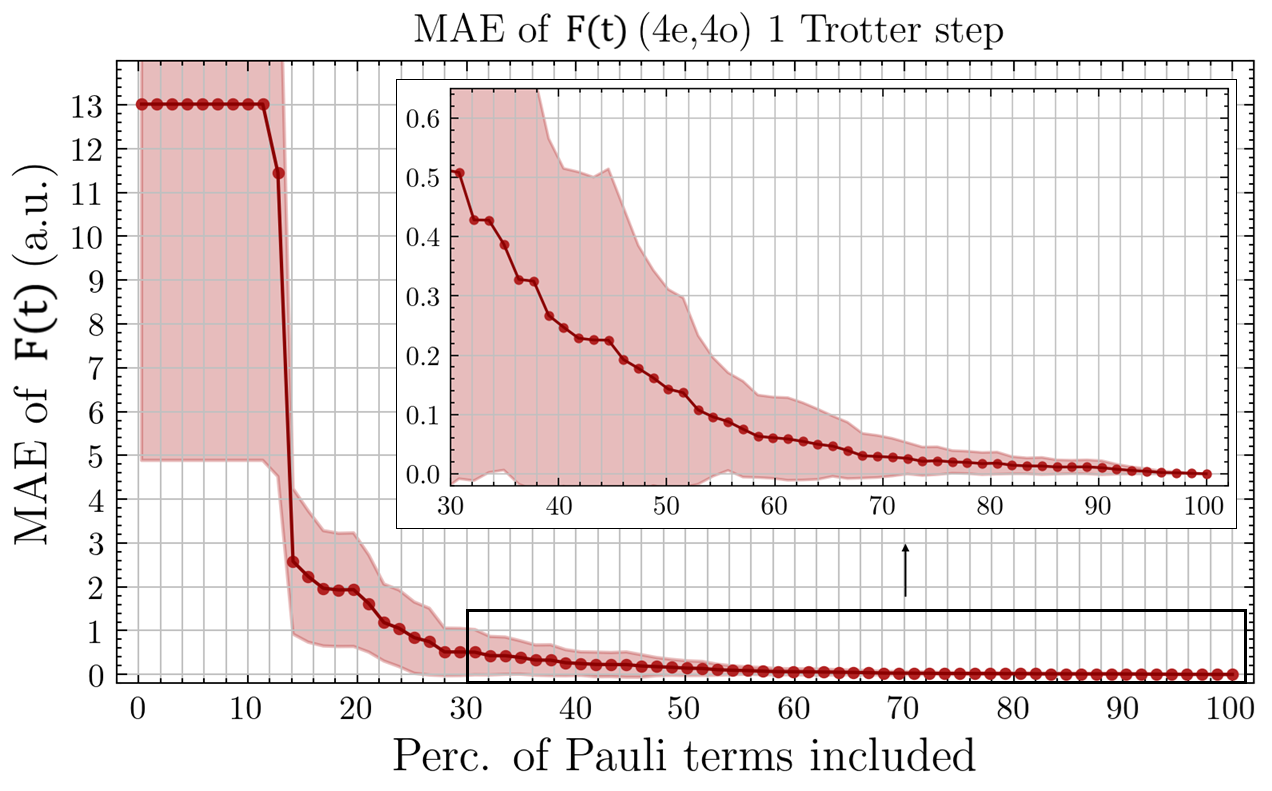}
    \caption{Mean absolute error (MAE) between trajectories emulated without noise over a range of terms left in the Hamiltonian and full Hamiltonian trajectory. The Hamiltonian and the initial state were taken for an exemplary molecule of active space 8 spin-orbitals, same as in Figure \ref{fig:time_evo_example_from_terms_included}. The dots are the means and the semi-transparent area is the standard deviation, both calculated from 50 equally spaced time steps on a range between 0 and 100 a.u.}
    \label{fig:time_evo_MAE_from_terms_included}
\end{figure}

To graphically show the influence of removing terms from the Hamiltonian in Figure \ref{fig:time_evo_example_from_terms_included}, the energy trajectories from different numbers of terms left in the Hamiltonian were plotted for 8 spin-orbitals case of an exemplary molecule. The trajectories were calculated on a noiseless simulator. Per each trajectory only the terms of the highest absolute values of coefficients were left in the Hamiltonian. The number of terms included in the given trajectory, shown as a percentage compared to all Hamiltonian terms, was color coded and marked on the colorbar. The mean absolute error (MAE) between each trajectory and trajectory based on the full Hamiltonian was shown in Figure \ref{fig:time_evo_MAE_from_terms_included}. For reference, for the observable error limit $\epsilon_{d(\langle \hat{A} \rangle )}=1\%$ for the max. time $100$ (a.u.), according to the presented truncation algorithm (Appendix~A) 42.45\% of terms should be left, and around 40\% we see the value of MAE close to 0.2, which is roughly 1\% of the smallest absolute values of the energy.

The final objective of our use case is not the trajectories themselves, but the reactivities of drugs predicted with ML model based on them. In Figure \ref{fig:ML_prediction}, the prediction of reactivities was evaluated on the ML model and dataset as described in the Methodology Section~\ref{sec:methodology}. Per each active space, the PLS model was trained for a number of components yielding the best evaluation scores for a given active space size. Training and testing was done with 5-fold cross-validation. The evaluation was performed for each active space over input trajectories simulated with truncated Hamiltonians from minimally 1\% of terms left to 100\% of terms left (full Hamiltonian). Each data point per active space per percentage of Hamiltonian terms left denotes a single training and testing model from scratch.  

As evaluation metric, the RMSE was chosen, to ensure comparability with previous paper QDDRD setup\cite{Montgomery2023}. In this application, the RMSE should be interpreted primarily as a task-level metric for preserving the predictive performance of the full-Hamiltonian quantum fingerprints, rather than as a universal chemical-accuracy threshold. Therefore, we regard a truncated simulation as useful when its RMSE remains comparable to the full-Hamiltonian baseline and does not lead to a significant degradation in $R^2$.  

For strong truncation with small percentages of terms left, there is an expected decrease in performance, both for RMSE and $R^2$. After considering the percentage threshold for $\epsilon_{d(\langle \hat{A} \rangle )}=1\%$ taken from reference molecule from the Result Section~\ref{subsec:circuits_reduction_results}, all truncations with more terms allow to maintain similar performance as for full Hamiltonian. That is also expected since the terms included above that threshold have small coefficients inducing small changes to trajectory. 


\subsection{Circuits reduction results}
\label{subsec:circuits_reduction_results}

\begin{figure}[h!]
    \centering
    \includegraphics[width=0.9\linewidth]{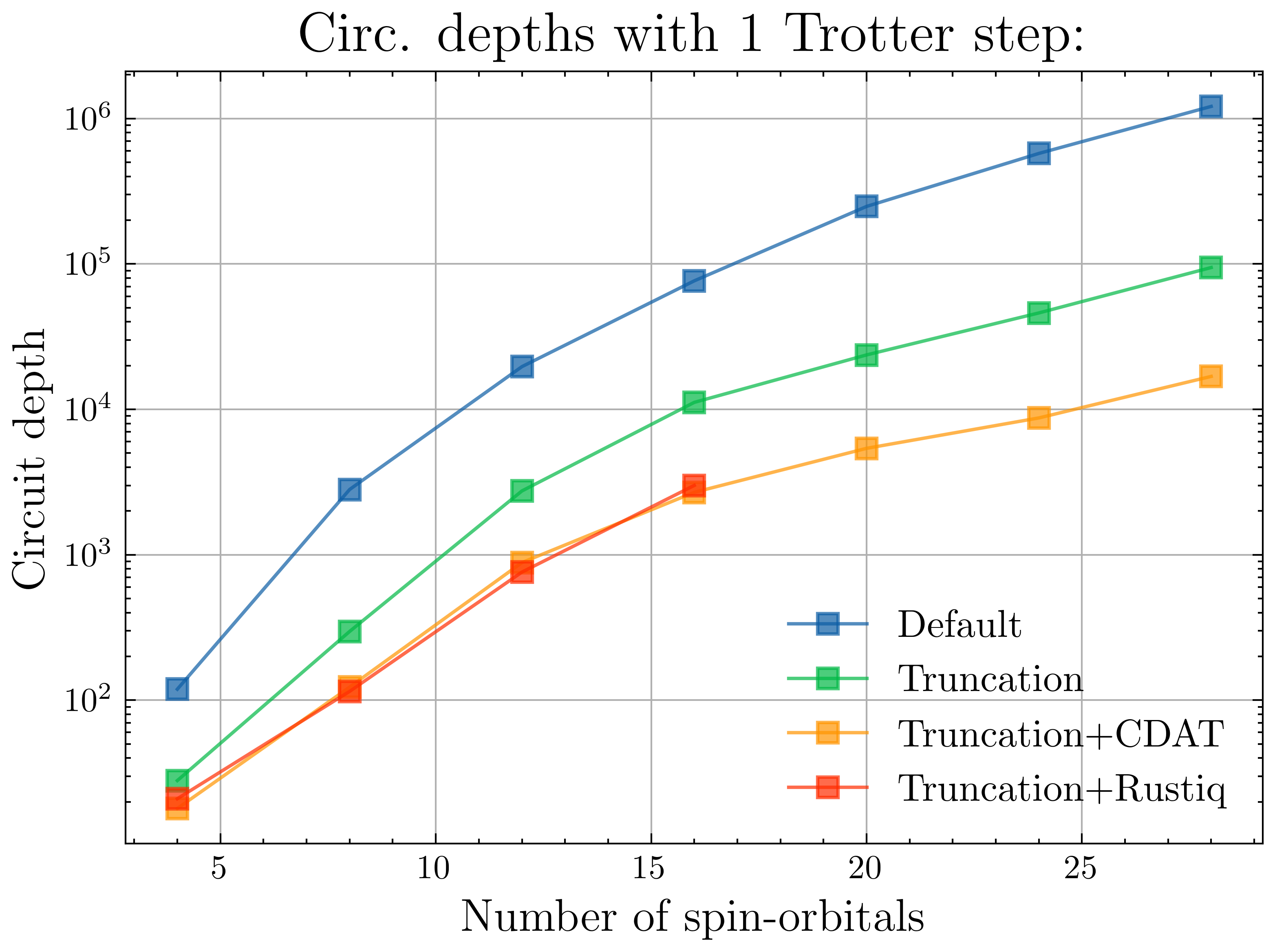}
    \caption{Depths of the quantum circuits for 1 Trotter step for time evolution simulation of qubit mapped effective fragment Hamiltonian for an exemplary molecule through active space size with Jordan-Wigner qubit mapping and the Hartree-Fock initial state. }
    \label{fig:Circuit depths reduction}
\end{figure}

A combination of Hamiltonian terms truncation reduction (for observable error limit $\epsilon_{d(\langle \hat{A} \rangle )}=1\%$ for the max. time $14$ (a.u.)) with CDAT or Rustiq reduces the circuit depths by slightly above one order of magnitude as shown in Figure \ref{fig:Circuit depths reduction} on an exemplary molecule. The same results up to 16 qubits numerically were shown in Table \ref{tab:circuit_metrics} along with the circuit depths for real QPU after transpilation to IBM's Heron R2 architecture. Circuits were transpiled with Qiskit 1.2\cite{Qiskit} generic pass manager with \textit{optimization\_level}=2 on all-to-all connectivity with $\textit{Clifford}+R_z$ set of gates (Figure \ref{fig:Circuit depths reduction} and left part of Table \ref{tab:circuit_metrics}) and IBM Heron R2 architecture (right part of Table \ref{tab:circuit_metrics}). The depths are the best results from 50 transpilation trials. All this data was plotted in the abstract figure. The displayed methods are `Default' denoting full Hamiltonian-based circuit (`No reduction'), `Truncation' - truncated Hamiltonian-based circuit (`Reduction level 1'), `Truncation+CDAT' - truncated Hamiltonian-based circuit with Clifford Decomposition and Transformation (CDAT) applied (`Reduction level 2'), and `Truncation+Rustiq' - truncated Hamiltonian-based circuit with Rustiq transpiler's circuit synthesis applied (`Reduction level 3').

\vspace{\baselineskip}
\noindent
\begin{minipage}{\linewidth}
    \captionsetup{type=figure} %
    \centering
    \subcaptionbox{}{\includegraphics[width=0.98\linewidth]{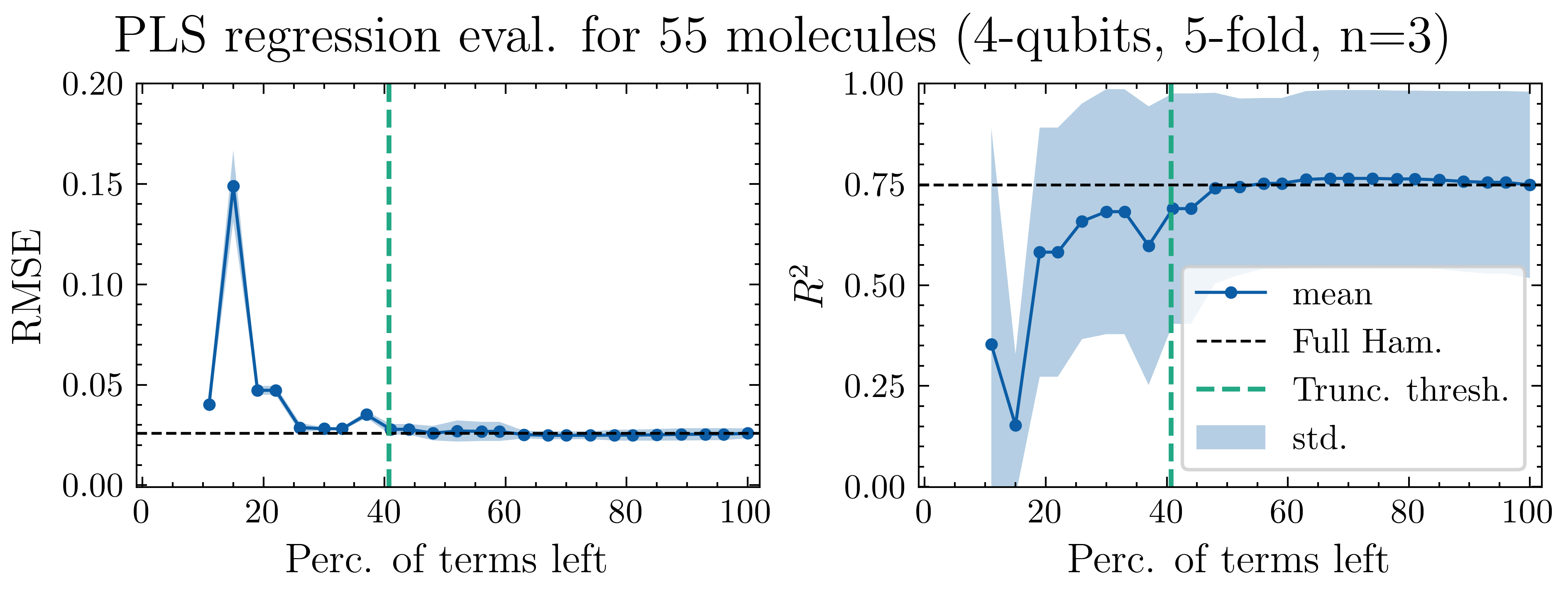}}
    \subcaptionbox{}{\includegraphics[width=0.98\linewidth]{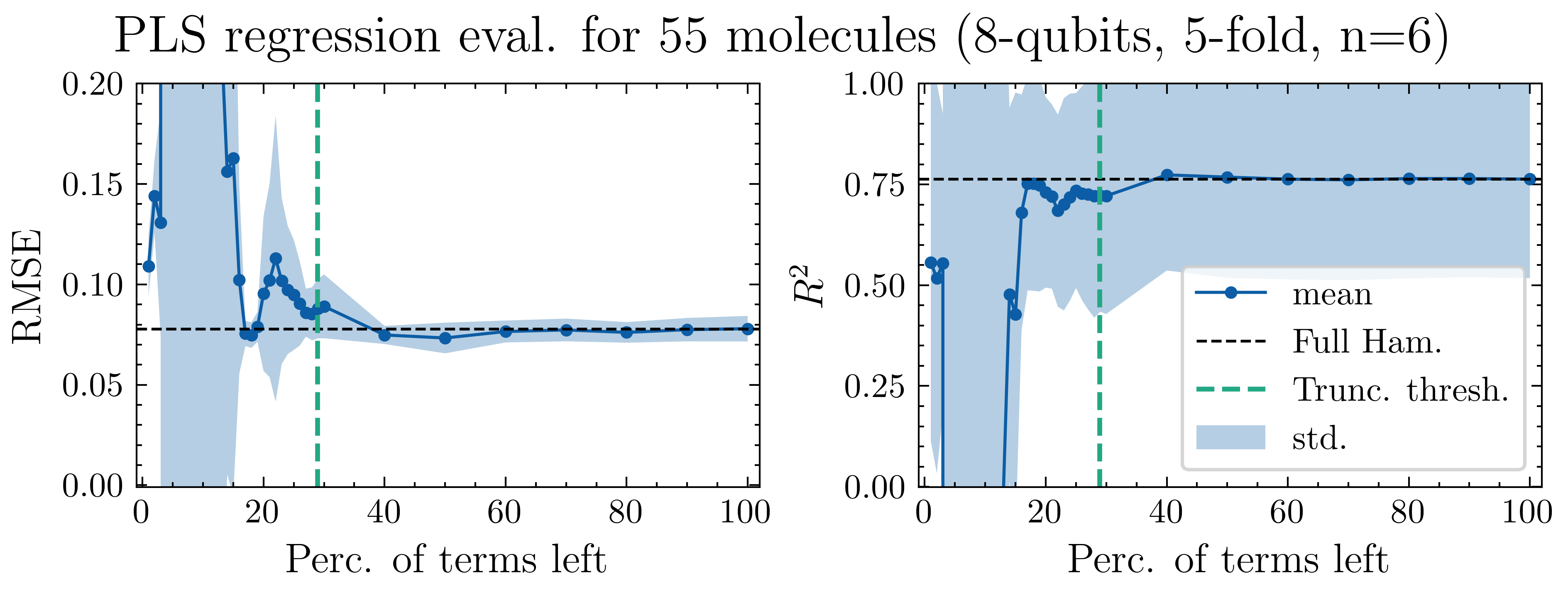}}
    \subcaptionbox{}{\includegraphics[width=0.98\linewidth]{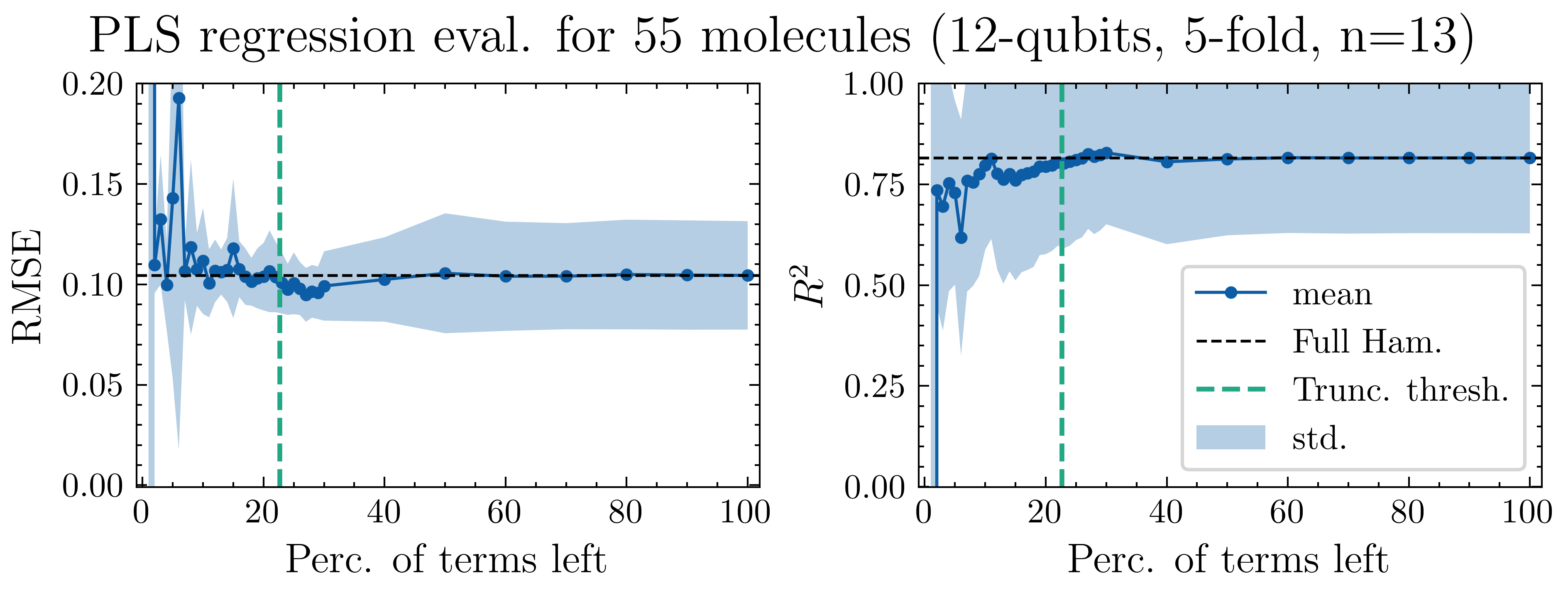}}
    \caption{Evaluation metrics comparison over a) (2e,4so), b) (4e,8so) and c) (6e,12so) active spaces for PLS. The left-hand side shows the Root Mean Squared Error (RMSE), and the right-hand side explained variance ($R^2$) from test sets (55 molecules). Numbers of PLS components are listed in the titles as `\textit{n}'. The horizontal line is the score for the full Hamiltonian (100\% of terms). For reference, the truncation threshold for $\epsilon_{d(\langle \hat{A} \rangle )}=1\%$ for the max. time $14$ (a.u.) time based on the molecule from the Result Section~\ref{subsec:circuits_reduction_results} was drawn vertically.}
    \label{fig:ML_prediction}
\end{minipage}
\vspace{\baselineskip}

\subsection{Real QPU results}
\label{subsec:real_qpu_results}

\begin{figure}[h!]
    \centering
    \includegraphics[width=0.98\linewidth]{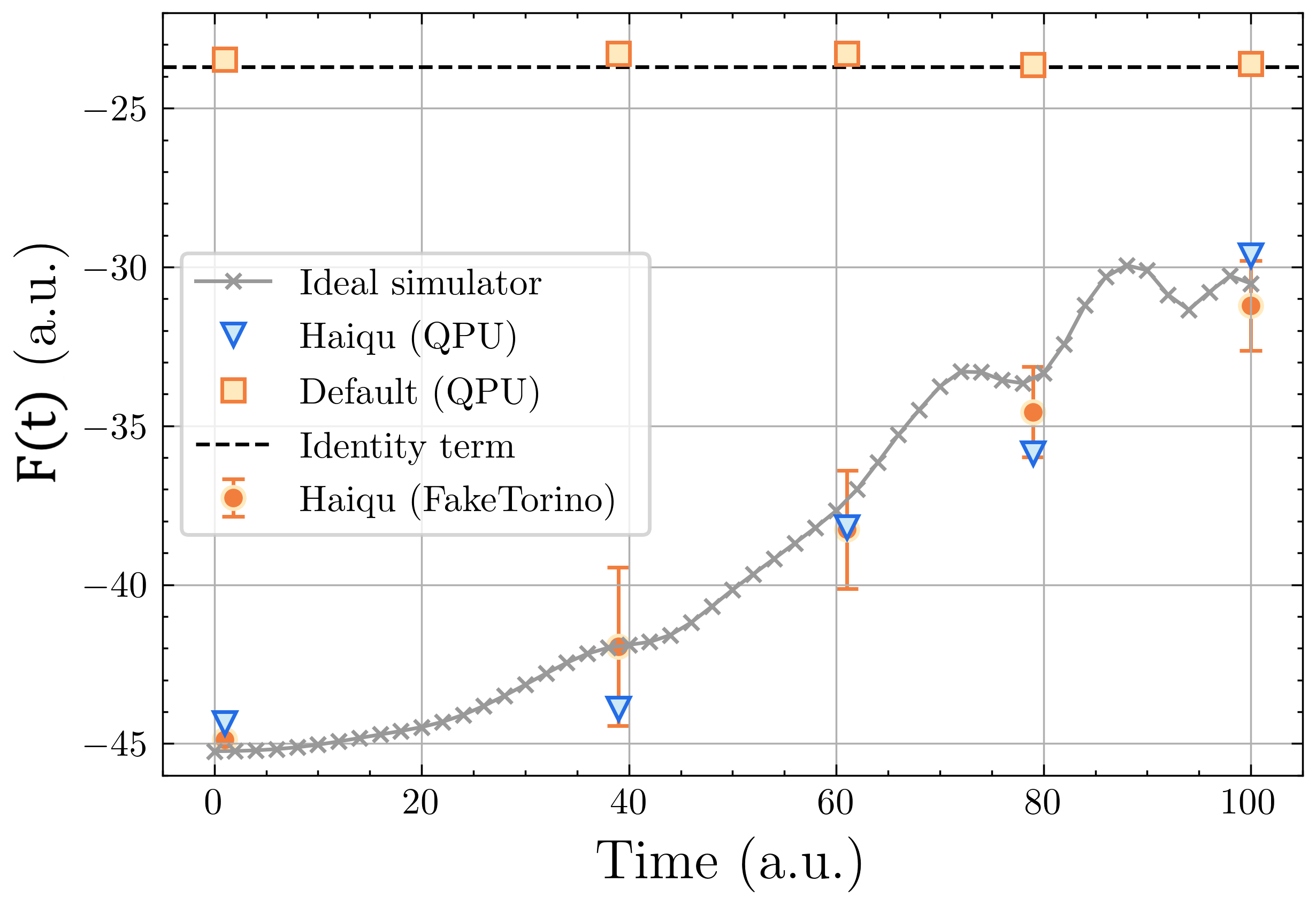}
    \caption{Real quantum computers results comparison for the run, IBM's Heron R2 156-qubits Marrakesh device, with Qiskit build-in error mitigation (QEM) and Haiqu middleware (Haiqu). Here for error bars estimation we run 10 executions of a similar setup on IBM's FakeTorino emulated QPU, based on Heron architecture. Time steps 1, 39 and 100 are executed with OE while for time steps 61 and 79 we combine OE with Operator backpropagation post-processing technique (see ~Appendix~B).}
    \label{fig:Heron QPU run with Haiqu middleware and in-house mitigation}
\end{figure}

Reference results from runs with generic error mitigation (as described in the Methodology Subsection~\ref{subsec:error_mitigation_methods}) and with Haiqu middleware, are shown in Figure \ref{fig:Heron QPU run with Haiqu middleware and in-house mitigation} for the (4e/8so) active space mapped on 8-qubits. Circuits were prepared with constrained truncation for observable error limit $\epsilon_{d(\langle \hat{A} \rangle )}=0.5\%$ for the max. time $14$ (a.u.), that is the maximum time step included in ML model input feature, and after CDAT. After transpilation to the IBM Heron R2 backend for the direct QPU baseline, the full ISA circuit, excluding additional error-mitigation and error-suppression gates, had depth 1330. Through middleware decomposition, while the largest transpiled sub-circuit had depth 216 and contained 113 two-qubit gates, after extension with additional gates required by the Optimized Execution process, the largest single quantum circuit block successfully executed on IBM Marrakesh had depth 371 and contained 216 two-qubit gates, representing 

\vspace{\baselineskip}
\noindent
\begin{minipage}{\linewidth}
    \captionsetup{type=table} 
    \centering
    \subcaptionbox{Quantum simulators circuits for all-to-all qubits connectivity for Clifford+$R_z$ gates}[0.88\linewidth]{\resizebox{\linewidth}{!}{
    \begin{tabular}{|c|c|c|c|c|}
        \hline
        & \multicolumn{4}{c|}{\textbf{Active Space Size}} \\ 
        \hline
        & \textbf{(2e/4so)} & \textbf{(4e/8so)} & \textbf{(6e/12so)} & \textbf{(8e/16so)} \\ 
        \hline
        \textbf{Default} & 119& 2812& 19669& 76523\\ 
        \hline
        \textbf{Truncation} & 28& 297& 2748& 11169\\
        \textbf{(reduction)}& ({4.3x})& ({9.5x})& ({7.2x})&({6.9x})\\ 
        \hline
        \textbf{CDAT} & 18& 123& 884& 2688\\
        \textbf{(reduction)}& (6.6x)& ({22.9x})& ({22.3x})&({28.5x})\\ 
        \hline
        \textbf{Rustiq} & 21& 115& 760& 2999\\ 
        \textbf{(reduction)}& ({5.7x})& ({24.5x})& ({25.9x})& ({25.5x})\\ 
        \hline
    \end{tabular}
    }}
    \quad
    \subcaptionbox{ISA circuits on IBM's Heron R2 architecture}[0.88\linewidth]{\resizebox{\linewidth}{!}{
    \begin{tabular}{|c|c|c|c|c|}
        \hline
        & \multicolumn{4}{c|}{\textbf{Active Space Size}} \\ 
        \hline
        & \textbf{(2e/4so)} & \textbf{(4e/8so)} & \textbf{(6e/12so)} & \textbf{(8e/16so)} \\ 
        \hline
        \textbf{Default} & {293}& {7015}& {49388}& {195611}\\ 
        \hline
        \textbf{Truncation} & {102}& {823}& {8104}& {33356}\\
        \textbf{(reduction)}& ({2.9x})& ({8.5x})& ({6.1x})&({5.9x})\\ 
        \hline
        \textbf{CDAT} & {39}& {585}& {3916}& {12650}\\
        \textbf{(reduction)}& ({7.5x})& ({12.0x})& ({12.6x})&({15.5x})\\ 
        \hline
        \textbf{Rustiq} & {57}& {677}& {6101}& {27140}\\ 
        \textbf{(reduction)}& ({5.2x})& ({10.4x})& ({8.1x})& ({7.2x})\\
        \hline
    \end{tabular}
    }}
    \caption{Pre-transpiled (a) and transpiled (b) circuit depths for different circuit reduction options on an exemplary molecule simulation circuit for 1 Trotter step. Truncation circuits were generated for limit $\epsilon_{d(\langle \hat{A} \rangle )}=1\%$ for the max. time $14$ (a.u.)}
    \label{tab:circuit_metrics}
\end{minipage}

\noindent
one of the largest electronic structure Hamiltonian dynamics simulations implemented on current quantum hardware in terms of circuit complexity.

\section{Discussion}
\label{sec:discussion}

The significant reduction Hamiltonian dynamics circuit through our comprehensive optimization framework represents the progress in electronic structure Hamiltonian simulation.While the full transpiled circuit still exceeded current hardware capabilities for direct execution, our implementation through middleware-enabled decomposition establishes a new benchmark for what is possible on NISQ devices. This achievement not only demonstrates the effectiveness of our circuit reduction techniques but also highlights the importance of sophisticated decomposition strategies for practical quantum chemistry calculations. 

The approach proposed in this work might be further extended with other methods to enable even larger active spaces runs. The emphasis could be put on further circuit depth reduction, using e.g. circuit knitting\cite{Piveteau2024}, repeated qubit tapering, methods including post-processing\cite{Robledo-Moreno2024, Clinton2021}, testing more qubit-mapping methods\cite{Setia2018}, particularly the ones considering qubit connectivity in real backends \cite{Miller2023,Chiew2023}, tensor network approximation techniques such as Approximate Quantum Compiler (AQC)\cite{madden2021bestapproximatequantumcompiling}, and even leveraging subspace propagation, e.g. time-dependent Krylov/Arnoldi dynamics~\cite{takahashi2025krylovsubspacemethodsquantum}.

Our ML model evaluation revealed that terms with larger absolute coefficient values have greater impact on both trajectory accuracy and predictive model performance than terms with smaller coefficients. The PLS model demonstrates robustness to truncation error, maintaining comparable performance when using Hamiltonians truncated up to the chosen error bound of $\epsilon_{d(\langle \hat{A} \rangle )}=1\%$ across tested active-space sizes. While the circuit-depth improvement from truncation alone is modest, the approach is straightforward to implement and appears scalable over active spaces. The remaining discrepancy between ideal and hardware results arises from several distinct error sources. In noiseless simulations, the dominant controlled approximation is Hamiltonian term truncation, whose effect is isolated in Figs.~\ref{fig:time_evo_example_from_terms_included}--\ref{fig:ML_prediction} through trajectory-level errors and downstream RMSE/$R^2$ changes. Separately, PF algorithmic error, would influence ML model accuracy, that was explored initially in the previous work\cite{Montgomery2023}.
In the QPU experiments, additional errors arise from hardware-aware transpilation and routing, finite sampling, device noise in deep two-qubit-gate layers, imperfect error mitigation, and the middleware decomposition/reconstruction procedure.

\section{Conclusions}
\label{sec:conclusions}
 
This work demonstrates that practical Hamiltonian dynamics relevant to covalent inhibitor reactivity can be executed on present‑day quantum hardware when the problem is framed and optimized end to end for execution rather than algorithmic ideality. By combining Hamiltonian term truncation, circuit‑level decomposition, and hardware‑aware transpilation with middleware‑enabled execution, we show that circuits far beyond the depth limits of direct execution can be deployed in practice. For an eight-qubit system, Hamiltonian dynamics circuits with transpiled ISA depth exceeding one thousand were rendered executable through controlled decomposition into sub-circuits compatible with current superconducting devices.

Crucially, the resulting quantum outputs retain sufficient information to support a downstream reactivity prediction task, indicating that aggressive circuit reduction does not necessarily preclude chemical practicality. Rather than optimizing Hamiltonian simulation fidelity in isolation, the results emphasize the importance of evaluating quantum workflows through task‑level performance metrics that reflect their intended use. 

The contributions of this work are therefore best understood as an execution case study that exposes concrete trade‑offs between circuit depth, approximation, and downstream value in the NISQ regime. While future improvements in hardware fidelity and connectivity will expand the accessible problem space, the need to reason explicitly about end‑to‑end executability will remain central to the design of practical quantum chemistry applications. Future work will focus on extending this execution‑oriented approach to larger active spaces and alternative chemical observables, while systematically evaluating how improvements in hardware fidelity, connectivity, and error mitigation shift the identified executability boundaries. 

\section*{Declarations}

\textbf{Supplementary information.}   
Supplementary Information is available free of charge.  

\textbf{Competing interests.}   
The authors declare that they have no competing interests.  

\textbf{Data availability.}   
The complete dataset generated and analyzed during the current study is not publicly available due to 8 molecules within, that were provided by GSK under a confidentiality agreement. The rest of the dataset, containing the molecules' PubChem IDs, their respective SMILES and 3D structures, is publicly available on GitHub repository\cite{kowalik2025github}. 

\vspace{-2.0em}
\noindent
\begin{figure}[h]
    \includegraphics[width=0.9\linewidth]{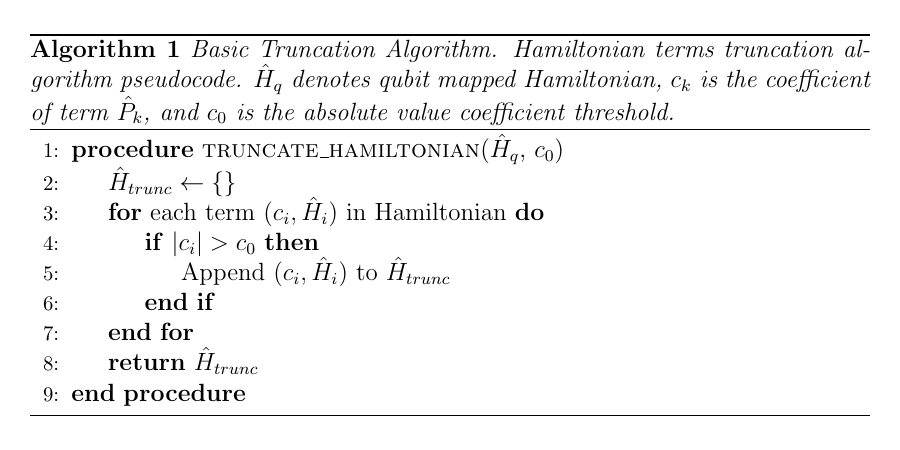}
    \caption{Pseudocode for a basic implementation of Hamiltonian terms truncation algorithm.}
    \label{alg:truncation alg. description}
\end{figure}
\vspace{-2.0em}

\noindent
Additional data supporting the findings of this study are also available upon reasonable request.  

\textbf{Author contributions.}   
SG, MK, HM, and RP conceived and designed the project, with SG leading the initial coordination. PL later assumed scientific leadership and overall coordination of the work. MK led the technical development and implementation, performing the majority of the computational work and analysis. MK, VP, and SM evaluated circuit reduction techniques, with VP and SM specifically focusing on Rustiq evaluation. MM, OH, and VL performed end-to-end experiments on real quantum backends, including implementation with and without middleware. PP provided the pharmaceutical dataset and valuable insights into pharmaceutical applications. MK and PL led the manuscript writing, with significant contributions from PP and MM. PL provided overall coordination of the work and guided the narrative direction, with additional support from RP and HM in shaping the structure and storyline.

\textbf{Acknowledgements.}  

Portions of the manuscript benefited from automated language‑assistance tools (Anthropic's Claude 3.5 and Microsoft Copilot) used for grammar and clarity checking; all scientific content, analysis, and conclusions are the sole responsibility of the authors.

\section*{Appendix A. Hamiltonian\ terms\ truncation}
\label{SI_sec:hamiltonian_terms_truncation}


A schematic representation of the truncation algorithm used in this work, through pseudocode is shown in Figure \ref{alg:truncation alg. description}.

The threshold $c_0=\epsilon_{d\langle\hat{A}\rangle}/(2dt)$ is used as a per-term truncation heuristic: an individual omitted term with $|c_i|<c_0$ is expected to have a limited first-order effect on the target observable. This does not constitute a rigorous aggregate error bound, since the cumulative worst-case error scales with $\sum_{i\in\mathrm{removed}}|c_i|$ and with $\|\hat{A}\|$. Thus, $\epsilon_{d\langle\hat{A}\rangle}$ should be interpreted as a relative tolerance with respect to $\|\hat{A}\|$, and the final truncation accuracy is assessed empirically.

\begin{figure*}
    \centering
    \includegraphics[width=0.9\linewidth]{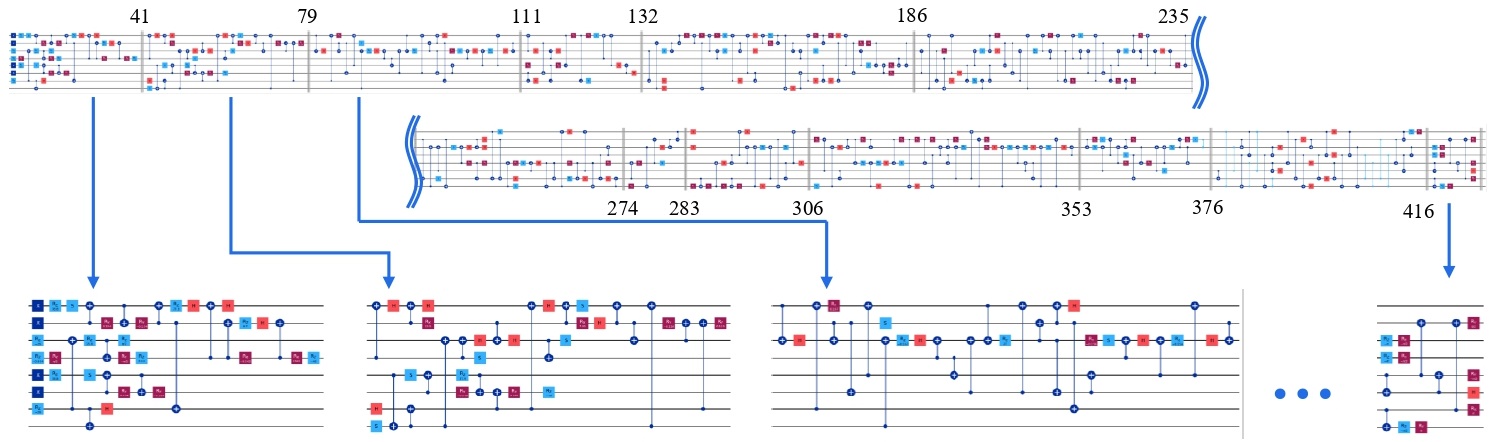}
    \caption{Decomposition of the 8-qubit logical circuit (similar circuit for the time points 1, 39, 61, 79 and 100) into 13 blocks with sizes ranging from 9 to 55 gates before transpilation to the device. Numbers indicate the cumulative logical gate count at which barriers are inserted. For time points 61 and 79 we execute circuit combining OE with Operator Backpropagation technique which post processes the second part of the circuit indicated with blue lines.}
    \label{fig:haiqu_circuit_decomposition}
    \vspace{-1em}
\end{figure*}

\section*{Appendix B. Detailed results of QPU runs with middleware}
\label{SI_sec:Haiqu_results}

In this section, all time steps are in atomic units, and the (a.u.) notation is omitted for readability.  

The partitioning strategy within Optimized Execution (OE) is based on multiple variables including circuit structure, layout mapping, sub-block entanglement, and fidelity of its execution on a noisy QPU. In the case of the quantum circuits used for the energy calculation at time steps 1, 39, 61, 79, and 100, the pre-transpilation logical circuit contains 431 gates and is partitioned into $N_{blocks}=13$ sub-blocks, as shown in Figure \ref{fig:haiqu_circuit_decomposition}. This 431-gate count refers to the logical circuit before hardware transpilation, whereas the depth 1330 reported in the main text refers to the corresponding full ISA circuit after transpilation to the IBM Heron R2 backend. Here, the barriers are inserted after logical gate indices 41, 79, 111, 132, 186, 235, 274, 283, 306, 353, 376, and 416. This produces logical sub-circuits with sizes ranging from 9 to 55 gates before transpilation to the device. After transpilation to the IBM Marrakesh backend, the largest sub-circuit submitted to the QPU had depth 216 and contained 113 two-qubit gates. After extension with additional gates required by the OE process, the largest executed quantum circuit block had depth 371 and contained 216 two-qubit gates.

With the OE, the execution of the full circuit on a QPU is performed as $N_{sampling} \times n_{blocks}$ QPU executions of the modified sub-block circuits, with an overhead of $N_{sampling}$ related to sampling the initial state and internal circuit structure, related to specifics of the method. For the problem circuits at hand $N_{sampling} = 600$. Each execution includes the application of custom noise mitigation and suppression pipeline involving dynamical decoupling, Pauli twirling, and readout error mitigation. With $N_{twirls}=8$ twirled circuit copies this sums up to $4800$ circuit executions per sub-block, however, we keep the overall shot budget for a twirled set of $8$ circuits $N_{shots}=10000$. We test our approach on noisy IBM devices using the Heron architecture - the actual IBM Marrakesh QPU, and the IBM FakeTorino simulator, which provides the layout and a high-fidelity noise model of a Heron architecture. 

Due to the limited hardware QPU accessibility and overhead of the method, part of the execution batches was run on the actual IBM Marrakesh QPU and some of the intermediate blocks are executed on noisy IBM FakeTorino simulator. The energy values estimation at time steps 1, 39, and 100 are

\noindent
\vspace{\baselineskip}
\begin{minipage}{\linewidth}
    \captionsetup{type=figure} 
    \centering
    \subcaptionbox{}{\includegraphics[width=0.6\linewidth]{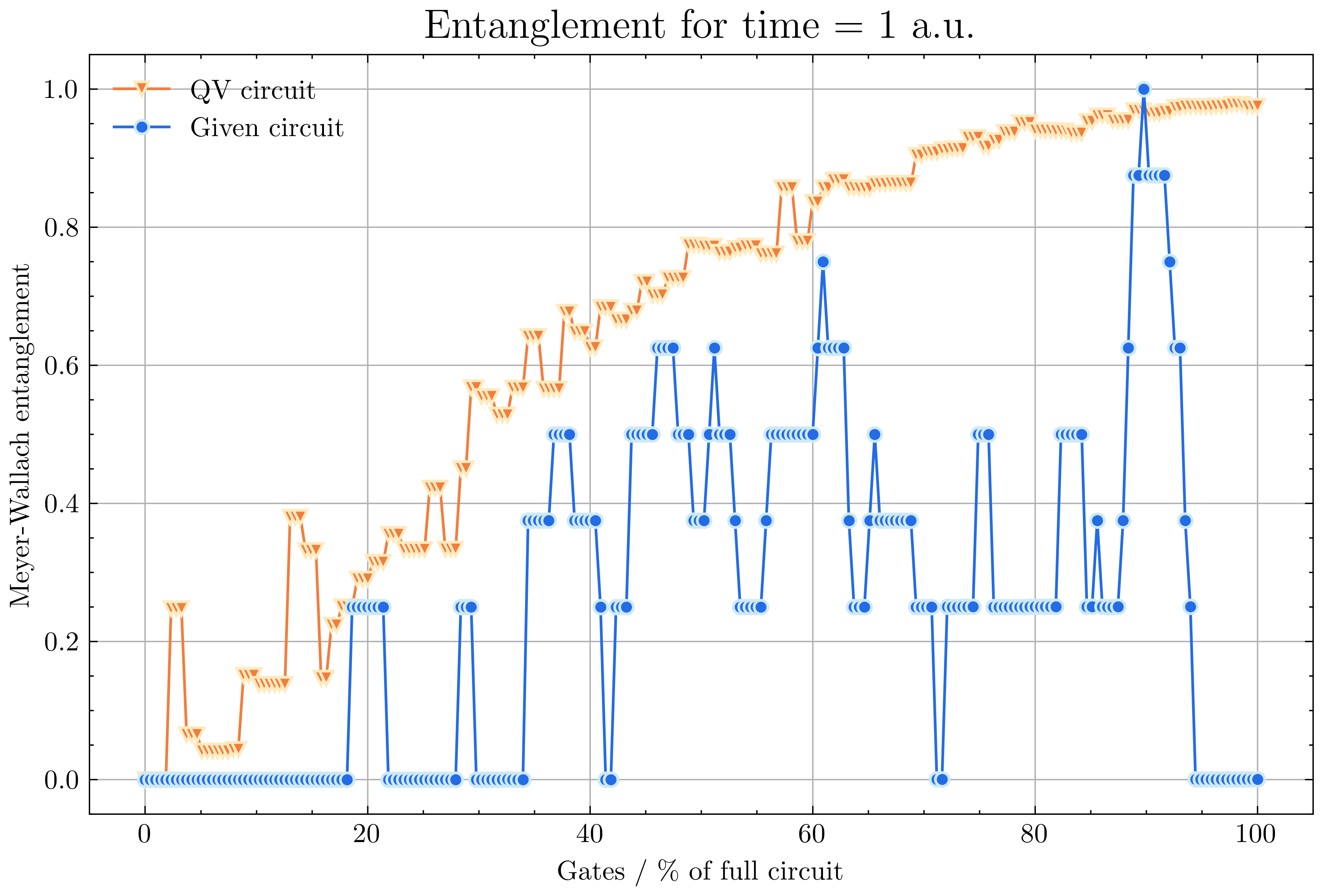}}
    \quad
    \subcaptionbox{}{\includegraphics[width=0.6\linewidth]{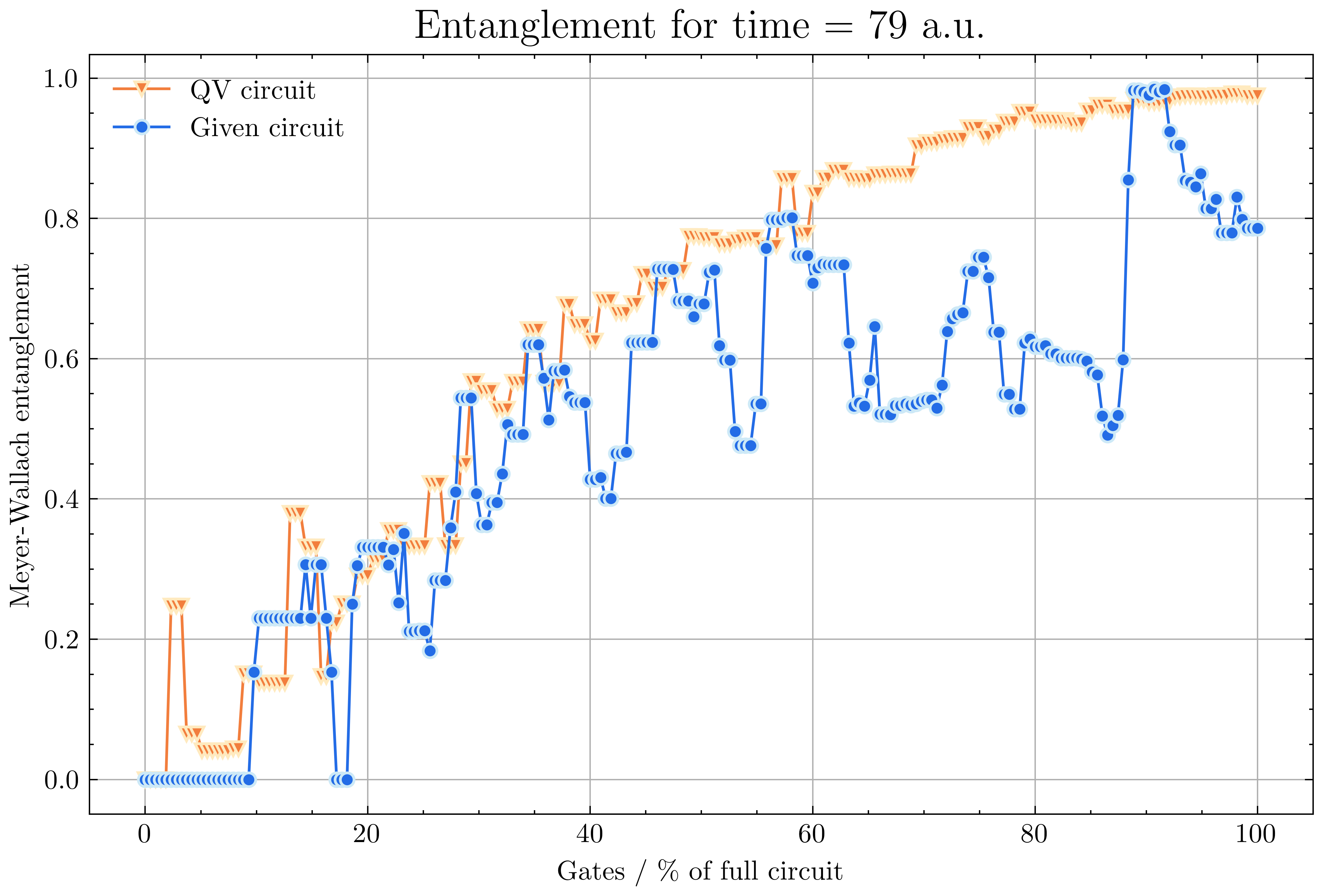}}
    \caption{Comparison of entanglement growth in the 8 qubit circuit at time steps 1 and 79 through Meyer-Wallach entanglement\cite{Brennen2003}. For intermediate time step 79 entanglement generated by evolution circuit (blue curve) quickly saturates to values comparable or higher than average random Quantum Volume circuit generates (orange curve).}
    \label{fig:haiqu_intermidiate_entanglement}
\end{minipage}

\noindent
obtained using such a combined approach. For time steps 61 and 79 as a proof of principle we combine the approach with Operator Backpropagation (OB)\cite{Fuller2025}. To this end, for time steps 61 and 79 the last part of the circuit indicated by curved blue dividers in Figure \ref{fig:haiqu_circuit_decomposition} is processed by OB. Here the application of OB is motivated due to significant generation of entanglement in the latter steps of the circuit at these particular time steps comparable to high entanglement generated by random Quantum Volume circuit of similar size (see Figure \ref{fig:haiqu_intermidiate_entanglement}). The initial observable was backpropagated with allowed approximation error of 0.001 per layer with unlimited number of qubit-wise commuting groups (exact backpropagation).

\bibliographystyle{IEEEtran}
\bibliography{IEEEabrv,main}


\end{document}